# Imaging the geometrical structure of complex symmetric molecules by above-threshold ionization spectrum in an IR+XUV laser field

Xu-Cong Zhou[1,2], Shang Shi[1,8], Fei Li[1,8], Ying-Chun Guo[3], Yu-Jun Yang[4], Qing-Tian Meng[2], Jing Chen[5,6], Xiao-Jun Liu[7] and Bingbing Wang[1,8,¯]

[1]*Laboratory of Optical Physics, Beijing National Laboratory for Condensed Matter Physics, Institute of Physics, Chinese Academy of Sciences, Beijing 100190, China*
[2]*School of Physics and Electronics, Shandong Normal University, Jinan 250358, China*
[3]*Department of Physics, School of Physics and Electronics, East China Normal University, Shanghai 200062, China*
[4]*Insititute of Atomic and Molecular Physics, Jilin University, Changchun 130012, China*
[5]*HEDPS, Center for Applied Physics and Technology, Peking University, Beijing 100871, China*
[6]*Institute of Applied Physics and Computational Mathematics, Beijing 100088, China*
[7]*State Key Laboratory of Magnetic Resonance and Atomic and Molecular Physics, Wuhan Institute of Physics and Mathematics, Chinese Academy of Sciences, Wuhan 430071, China*
[8]*University of Chinese Academy of Sciences, Beijing 100049, China*

Based on the frequency-domain theory, we put forward a method of imaging the complex symmetrical polyatomic molecules by using IR+XUV two-color laser fields. Although the wave function of a polyatomic molecule, such as $SF_6$, is quite complex in momentum space, it can be simplified by several terms in the region of high value of momentum, where the destructive interference fringes from these terms carry the information of the molecular structure. In this work, we find that, since the XUV laser field may increases the kinetic energy of the

---

¯wbb@aphy.iphy.ac.cn

photoelectron dramatically in the above-threshold ionization (ATI) process of a molecule in an IR+XUV two-color laser field, the information of the molecular structure can be obtained by analyzing the interference fringes of the ATI spectrum in the high-energy region, where the simplified wave function of the molecule plays a dominant role in this part of the ATI spectrum. This method provides a unique accessible route towards imaging polyatomic molecules from a frequency perspective.

## I. INTRODUCTION

The emergence of strong laser fields has revealed many new nonlinear physical phenomena. One of the most important processes is above threshold ionization (ATI) [1-3], which was first found in forty years ago. As a direct photoionization process, ATI spectrum has been widely used in deriving the information of the molecular structures [4-5] and observing the dynamic processes of molecules in laser fields [6-7]. Later on, re-collision processes [8] in strong laser field, such as high-order above-threshold ionization (HATI) process [9-11] and high harmonic generation (HHG) [12-13], were observed. Therefore, in the re-collision process, the ionized electron driven by the reverse laser field has a certain probability to re-collide with the nuclei, so as to form fringes by interference between different re-collision paths carrying related molecular structure information in the electron energy spectrum [14-17] or harmonic spectrum [18-19]. However, for heteronuclear diatomic molecules [20], the ionized electron can only re-collide with the nucleus carrying positive charge, hence it is hard to accurately image the corresponding molecular structure by these re-collision spectra.

Recently, with the development of high frequency laser technique by high-order harmonic generation and free-electron laser, the pump-probe method becomes an effective tool to measure and control the ultrafast dynamic processes in the laser-matter interactions [21-27]. Especially, the ionization and dissociation of molecules can be observed and controlled by changing the laser conditions and the delay time between the two laser pulses [22-27]. For example, the lifetime of the charge-transfer processes in Argon dimers has been obtained by XUV+IR pump-probe experiments [24]. More recently, He *et al*. [21] demonstrated that the orientation- and molecular-orbital-resolved ionization can be used to retrieve the complex structure of molecules by controlling the intensity of two-color laser fields. However, since many electronic orbitals may participate in the pump-probe dynamic processes for complex molecules, it is still a tough task to derive the information of the molecular structure by analyzing the experimental data [28].

What discussed above is mainly based on time-domain investigation in coordinate space, and they are time consuming even if it can reflect the molecular structure information to some extent. In this work, we employ the frequency-domain theory based on a non-perturbative quantum electrodynamics, which has previously been used to handle problems related to strong fields [29-33], to deal with the angle-resolved ATI spectra of complex molecules in linearly polarized IR+XUV two-color laser fields. The purpose of using the two-color laser fields is to get a broad energy spectrum, where IR laser can broaden the spectrum and XUV laser can help us obtain high ionization probability of the electron in the molecule, which is beneficial to extract the complex molecular structural information. The research object here is the complex polyatomic molecule with high symmetry, *i.e.*, $SF_6$. We find that the molecular states in momentum space can be simplified in high energy region, and many interference formulas carrying the information of the bond lengths may be obtained by these simplified states. Therefore, the interference fringes in ATI spectrum can be used to predict the bond lengths by the corresponding interference formulas.

## II. THEORETICAL METHOD

In this work, we consider the ATI process of one polyatomic molecule in IR+XUV two-color laser fields. The frequency-domain representation of the method for ATI in two-color laser fields is given in [32-33]. Here this method is briefly summarized and applied for polyatomic molecule system. Based on the frequency-domain theory, the

Hamiltonian of the molecule-laser system can be written as

$$H = H_0 + U + V ,\qquad (1)$$

where $H_0 = \frac{(-i\nabla)^2}{2m_e} + \omega_1 N_{a_1} + \omega_2 N_{a_2}$ represents the energy operator for a free electron-photon system, $N_{a_i} = \frac{1}{2}(a_i^+ a_i + a_i a_i^+)$ is the photon number operator of the laser field with frequency $\omega_i$, where $a_i$ and $a_i^+$ are the annihilation operator and creation operator of the photon mode for $i=1, 2$. $U$ represents the molecular binding potential, and $V$ is the electron-photon interaction potential expressed by

$$\begin{aligned}V = &-\frac{e}{2m_e}\{(-i\nabla)\cdot[\mathbf{A}_1(\mathbf{r})+\mathbf{A}_2(\mathbf{r})]+[\mathbf{A}_1(\mathbf{r})+\mathbf{A}_2(\mathbf{r})]\cdot(-i\nabla)\}\\ &+\frac{e^2}{2m_e}[\mathbf{A}_1(\mathbf{r})^2+\mathbf{A}_2(\mathbf{r})^2]\end{aligned},\qquad (2)$$

where $\mathbf{A}_s(\mathbf{r}) = g_s(\hat{\varepsilon}_s e^{i\mathbf{k}\cdot\mathbf{r}} a_s + c.c.)$ is the vector potential with $g_s = (2\omega_s V_\gamma)^{-1/2}$ $(s=1,2)$ and $V_\gamma$ is the normalized volume of the laser field. The polarization vector of the laser field is defined as $\hat{\varepsilon} = \hat{\varepsilon}_x \cos(\xi/2) + i\hat{\varepsilon}_y \sin(\xi/2)$, where $\xi$ determines the polarization degree of the laser field, such as $\xi = \pi/2$ corresponds to circular polarization and $\xi = 0$ corresponds to linear polarization. The initial quantum state is expressed as $|\psi_i\rangle = \Phi_i(r)\otimes|l_1\rangle\otimes|l_2\rangle$, the eigenstate of the Hamiltonian $H_0+U$ with the associated energy $E_i = -E_B + (l_1+\frac{1}{2})\omega_1 + (l_2+\frac{1}{2})\omega_2$, in which $\Phi_i(r)$ is the ground state wave function of the molecule with the binding energy $E_B$, $|l_1\rangle$ and $|l_2\rangle$ are the Fock states of the two laser modes. The final state of the system can be denoted by $|\psi_f\rangle = |\psi_{\mathbf{P}_f m_1 m_2}\rangle$,

which is the quantized-field Volkov state in two-color laser fields that can be expressed as

$$\left|\psi_{\mathbf{P}_f m_1 m_2}\right\rangle = V_e^{-1/2} \exp[i(\mathbf{P}_f + u_{p_1}\mathbf{k_1} + u_{p_2}\mathbf{k_2})\cdot\mathbf{r}] \times \sum_{\substack{q_1=-m_1\\q_2=-m_2}}^{\infty} \Im_{q_1,q_2}(\zeta)^* $$
$$\times \exp\{-i[q_1(\mathbf{k}_1\cdot\mathbf{r}+\phi_1)+q_2(\mathbf{k}_2\cdot\mathbf{r}+\phi_2)]\}\left|m_1+q_1, m_2+q_2\right\rangle \quad (3)$$

where $V_e$ is the normalized volume, $\mathbf{p}_f$ is the final state momentum of the ionized electron, and $u_{p_1} = U_{p_1}/\omega_1$ ($u_{p_2} = U_{p_2}/\omega_2$), with $U_{p_1}$ ($U_{p_2}$) being the ponderomotive potential in the laser field. $\mathbf{k}_1(\mathbf{k}_2)$ denotes photon momentum of the laser field. The total energy of final state is

$$E_{\mathbf{P}_f m_1 m_2} = \frac{\mathbf{P}_f^2}{2} + (m_1+\frac{1}{2})\omega_1 + (m_2+\frac{1}{2})\omega_2 + u_{p_1}\omega_1 + u_{p_2}\omega_2, \quad (4)$$

where $m_1(m_2)$ is the photon number in the laser field. The generalized Bessel function $\Im_{q_1,q_2}(\zeta)$ can be written as

$$\Im_{q_1,q_2}(\zeta) = \sum_{q_3 q_4 q_5 q_6} J_{-q_1+2q_3+q_5+q_6}(\zeta_1) J_{-q_2+2q_4+q_5-q_6}(\zeta_2) $$
$$J_{-q_3}(\zeta_3) J_{-q_4}(\zeta_4) J_{-q_5}(\zeta_5) J_{-q_6}(\zeta_6) \quad (5)$$

with $\zeta \equiv (\zeta_1, \zeta_2, \zeta_3, \zeta_4, \zeta_5, \zeta_6)$,

$$\zeta_1 = 2\sqrt{\frac{u_{p_1}}{\omega_1}}\left|\mathbf{p}_f\cdot\hat{\varepsilon}_1\right|, \zeta_2 = 2\sqrt{\frac{u_{p_2}}{\omega_2}}\left|\mathbf{p}_f\cdot\hat{\varepsilon}_2\right|,$$

$$\zeta_3 = \frac{1}{2}u_{p_1}\cos\xi_1, \zeta_4 = \frac{1}{2}u_{p_2}\cos\xi_2,$$

$$\zeta_5 = 2\frac{\sqrt{u_{p_1}u_{p_2}\omega_1\omega_2}}{\omega_1+\omega_2}, \zeta_6 = 2\frac{\sqrt{u_{p_1}u_{p_2}\omega_1\omega_2}}{\omega_1-\omega_2}. \quad (6)$$

Now, the transition matrix element of ATI process can be written as

$$T_d = \left\langle\psi_f|V|\psi_i\right\rangle = V_e^{-1/2}[(u_{p_1}-q_1)\omega_1+(u_{p_2}-q_2)\omega_2]\Im_{q_1,q_2}(\zeta_f)\Phi_i(\mathbf{P}), \quad (7)$$

with $q_1 = l_1 - m_1$ ($q_2 = l_2 - m_2$) being the number of photons the electrons absorbing from two-color laser fields. And $\Phi_i(\mathbf{P})$ is the Fourier transform of the initial wave function $\Phi_i(r)$.

## III. NUMERICAL RESULTS

We may find from Eq. (7) that the transition matrix element is proportional to the wave function, *i.e.*, $T_d \propto \Phi_i(\mathbf{P})$, which indicates that the structure of the wave function can be mapped on the ATI spectrum. Therefore, the molecular structure information can be obtained by analyzing the ATI spectrum as these wave functions carrying the information, where the ATI spectrum satisfies some suitable conditions. In this section, we demonstrate that the two-color laser fields can provide such suitable ATI spectra to realize imaging molecules.

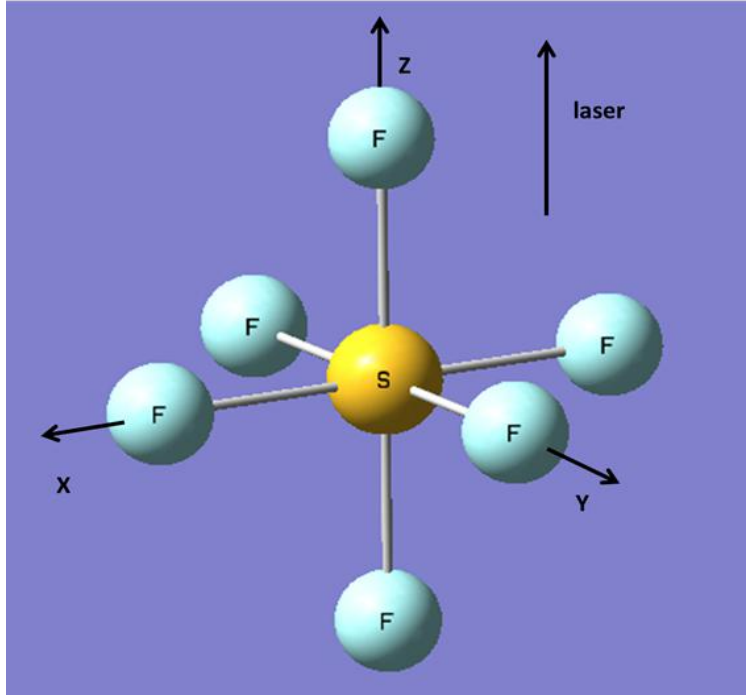

FIG 1. The skeleton of $SF_6$ molecules. The vector potentials of the two-color laser fields are along Z axis.

We now consider the ATI processes of molecule $SF_6$ in two-color laser fields. The ionization potential of $SF_6$ is 15.7eV [34]. Figure 1

shows the skeleton of the molecule SF$_6$, the origin of the coordinate system is set at the S atom. Atomic units are used throughout unless stated otherwise.

By using HF/6-311G$^*$ method [35-36] implemented by the Molpro software [37], the highest occupied molecular orbital (HOMO) of the SF$_6$ molecule includes three degenerated orbitals, which can be denoted as $\psi_1$, $\psi_2$ and $\psi_3$, according to their own symmetries. Here, the standard S-F bond length used in the calculation is $R_{SF}$=2.929 a.u..

According to the formula of Fourier transformation

$$\varphi(\mathbf{p}) = \frac{1}{(2\pi\hbar)^{3/2}} \int \psi(\mathbf{r}) e^{-i\mathbf{p}\cdot\mathbf{r}/\hbar} d^3 r , \tag{8}$$

we can obtain the wave functions for the three HOMO orbitals in momentum space. Firstly, we analyze the case of the SF$_6$ when the symmetry is $\psi_1$. The expression of the HOMO wave function $\psi_1$ in momentum space is Eq. (A4) in the Appendix A, and its density distribution in coordinate space is shown in Fig. A1.

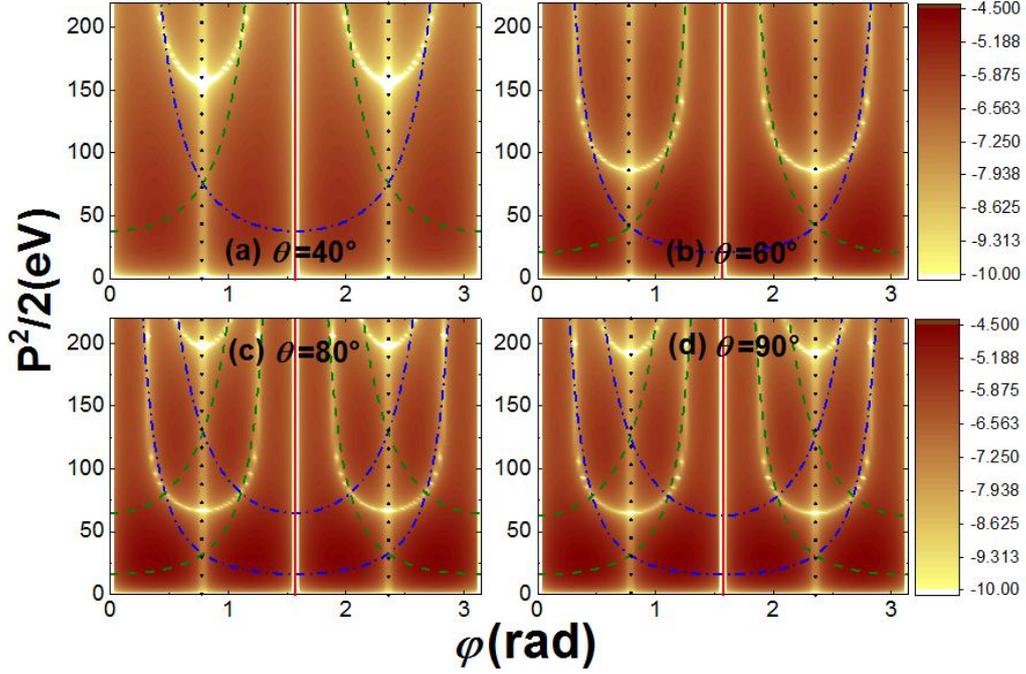

FIG. 2. The density distribution of $\psi_1$ wave function in momentum space for molecule $SF_6$.

Figure 2 shows density distribution of the wave function $\psi_1$ of $SF_6$ in momentum space with different zenith angles, where the horizontal axis represents azimuthal angle and vertical axis represents $p^2/2$ dependent on momentum. One may clearly find that there are many fringes in the density distribution of the wave function as shown in Fig. 2. By analyzing Eq. (A4), we may find that these fringes can be classified by three types: (1) when $\varphi=n\pi/2$ with n=1,2,…, the amplitude of the wave function is zero because of the dependence of each term in Eq. (A4) on the momentum $p_x$ and $p_y$, which is denoted by the red solid line in Fig. 2; (2) when $\varphi=n\pi/2+\pi/4$ with n=0,1,…, i.e., $p_x=\pm p_y$, the interference between each two corresponding terms in Eq. (A4) provides the destructive fringes in the density distribution of the wave function, as shown by the black dotted lines in Fig. 2; (3) Besides above fringes, we may find there exist more fringes with the energy increasing, as shown in Fig. 2. In fact, when the value of the energy is large enough, where the exponent part in each term of Eq. (A4) affects dramatically its value, the wave function $\psi_1$ can be approximately expressed by two terms, i.e.,

$$\psi_1 \approx -2.74966 \times 10^5 p_y \sin(R_{SF} p_x) e^{-\frac{p^2}{221.764}} \\ + 2.74966 \times 10^5 p_x \sin(R_{SF} p_y) e^{-\frac{p^2}{221.764}}, \tag{9}$$

hence, there will occur destructive fringes as one of the two terms becomes zero, which provide two fringes formula:

$$E = 0.5(n\pi)^2 / (\sin^2 \theta_f \cos^2 \varphi_f R^2), \tag{10}$$

and

$$E = 0.5(n\pi)^2 / (\sin^2 \theta_f \sin^2 \varphi_f R^2), \tag{11}$$

where $R$ is the S-F bond length. The green dashed and blue dash-dot lines in Fig. 2 are predicted by Eq. (10) and (11), respectively. We should notice that these lines for third types of fringes agree well with that of numerical results as the energy is larger than 150 eV. However, these fringes are blurred in the low energy region. This is because that more terms in the wave function in Eq. (A4) play an important role in the distribution in low energy region and their contributions destroy these fringe structures.

From the above analysis, we may find that the third type of the interference fringes in the density distribution of the wave function carries the information of the molecular structure, especially from the fringes of the distribution in high energy region. Therefore, according to the characteristic of the ATI spectrum predicted by Eq. (9), on which the density distribution of the molecular wave function maps, it is necessary to increase the energy of the ionized electron in order to increase the high energy part of the ATI spectrum. Here, we realize this idea by using IR+XUV two-color laser fields.

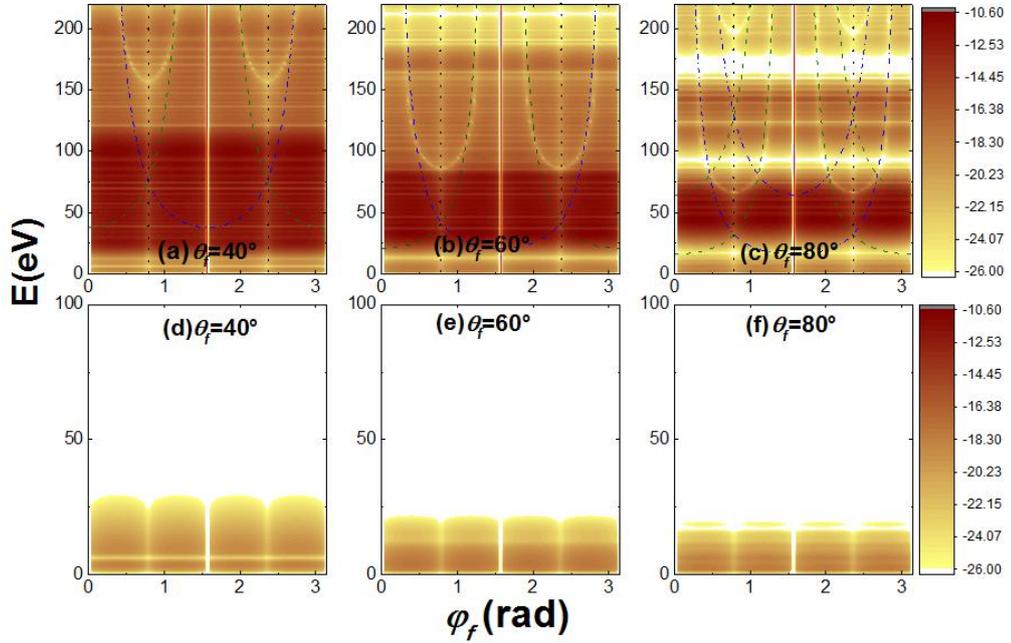

FIG. 3. The angle-resolved ATI spectra of $SF_6$ for initial state $\psi_1$ in the IR+XUV two-color fields (a-c) and monochromatic IR laser field (d-f) at $\theta_f = 40°$ (a and d), $60°$ (b and e) and $80°$ (c and f).

According to the interference formulas of Eq. (10) and (11), we can find that the energy increases dramatically as the angle between the direction of the laser polarization and the electron momentum direction decreases towards zero, hence the angle is chosen as $\theta_f \in (20°, 90°)$. In this paper, the polarization of the two-color linearly polarized laser field is along $z$ axis, as shown in Fig. 1. The intensities of the two laser fields are $I_1=I_2=1.0\times10^{14}$ W/cm², and the frequency $\omega_1$=1.55 eV and $\omega_2$=50$\omega_1$. The initial phases of the two-color laser fields are $\phi_1=\phi_2=0°$ for simplicity. Figure 3 (a-c) shows the ATI spectra in IR+XUV two color laser fields for initial wave function $\psi_1$ of SF$_6$ with the angle between the momentum direction of the electron and the polarization direction of the IR laser $\theta_f=40°$ (a), 60° (b) and 80° (c). For comparison, we also present the ATI spectrum by IR laser field in Fig. 3 (d-f) with the laser frequency $\omega$=1.55 eV and intensity $I$=1.0×10$^{14}$ W/cm² for $\theta_f=40°$ (d), 60° (e) and 80° (f). One may find that, the energy region of the ATI spectra for the monochromatic IR laser case is below 50eV, while the energy region of the ATI spectrum is enlarged obviously by using the IR+XUV laser fields as shown in Fig. 3. This result is attributed to that the participation of XUV laser improves the ionization probability, and the IR laser field accelerates the photoelectron during the ATI process [32]. In addition, the ATI spectra present multi-plateau structure, where the width of each plateau in the ATI spectrum decreases as the angle $\theta_f$ increases. These spectrum structures are determined by the energy conservation during the ionization process [33], where the cutoff of each plateau decreases with the angle increasing. More interestingly, we may find that the interference fringes emerge in the high energy region of these spectra, which are attributed to the interference fringes in the density distribution of the molecular wave function shown in Fig. 2. The green dashed and blue dash-dot lines are also from the interference formula of Eqs. (10) and (11), where the high energy part of these lines agrees well with the numerical results shown in Fig. 3. This result indicates that the analysis of the wave function which can be simplified by two terms in high energy region is suitable.

From Eqs. (10) and (11), we can derive the formulas about molecular bond length, which is

$$R = n\pi / (\sqrt{2E} |\sin\theta_f||\cos\varphi_f|), \tag{12}$$

$$R = n\pi / (\sqrt{2E} |\sin\theta_f||\sin\varphi_f|). \tag{13}$$

Therefore, by using the data of the ATI spectra, we can predict the S-F bond length of $SF_6$ by Eqs. (12) and (13), which are shown in Tables 1 and 2. The error in both tables is defined as $\Delta = (R_{avg} - R_{SF})/R_{SF}$. We can find that the error is smaller than 8% in both cases.

Tab. 1. The length of S-F bond length in $SF_6$ molecule by Eq. (12), where the data are taken from the ATI spectra shown in Fig. 3 for the molecular wave function of $\psi_1$.

| $\theta_f$ | $\varphi_f$ /rad | E/eV | R/a.u. | $R_{avg}$/a.u | Error $\Delta$ |
|---|---|---|---|---|---|
| **40°** | 1.12 | 190 | 3.00 | 3.03 | 3.45% |
| | 2.02 | 190 | 3.01 | | |
| | 1.15 | 210 | 3.04 | | |
| | 1.99 | 210 | 3.06 | | |
| **60°** | 1.26 | 190 | 3.17 | 3.10 | 5.84% |
| | 1.88 | 190 | 3.19 | | |
| | 1.26 | 210 | 3.02 | | |

| | 1.88 | 210 | 3.03 | | |
|---|---|---|---|---|---|
| **80°** | 1.26 | 190 | 2.79 | 2.73 | 6.79% |
| | 1.88 | 190 | 2.80 | | |
| | 1.26 | 210 | 2.65 | | |
| | 1.88 | 210 | 2.67 | | |

Tab. 2. The length of S-F bond length in $SF_6$ molecule by Eq. (13), where the data are taken from the ATI spectra shown in Fig. 3 for the molecular wave function of $\psi_1$.

| $\theta_f$ | $\varphi_f$ /rad | E /eV | R/a.u. | $R_{avg}$/a.u | Error$\Delta$ |
|---|---|---|---|---|---|
| **40°** | 0.45 | 190 | 3.01 | 3.03 | 3.45% |
| | 2.69 | 190 | 3.00 | | |
| | 0.42 | 210 | 3.05 | | |
| | 2.72 | 210 | 3.04 | | |
| **60°** | 0.31 | 190 | 3.18 | 3.10 | 5.84% |
| | 2.83 | 190 | 3.17 | | |
| | 0.31 | 210 | 3.03 | | |
| | 2.83 | 210 | 3.01 | | |
| **80°** | 0.31 | 190 | 2.80 | 2.72 | 7.14% |
| | 2.83 | 190 | 2.78 | | |
| | 0.31 | 210 | 2.66 | | |
| | 2.83 | 210 | 2.65 | | |

From the above discussion we can see that the wave function of $\psi_1$ in momentum space can be simplified by two main effective terms in Eq. (9). We now trace back to the coordinate space wave function, which corresponds to four main effective terms expressed as

$$\psi_1' = \frac{1}{\sqrt{2}}\{1.78005 y \exp\{-55.4441[(x-2.929)^2 + y^2 + z^2]\} \\ -1.78005 y \exp\{-55.4441[(x+2.929)^2 + y^2 + z^2]\} \\ -1.78005 x \exp\{-55.4441[x^2 + (y-2.929)^2 + z^2]\} \\ +1.78005 x \exp\{-55.4441[x^2 + (y+2.929)^2 + z^2]\}\} \quad (14)$$

We find that these four terms are from four F atoms in XY plate perpendicular to the direction of the laser field as shown in Fig.4.

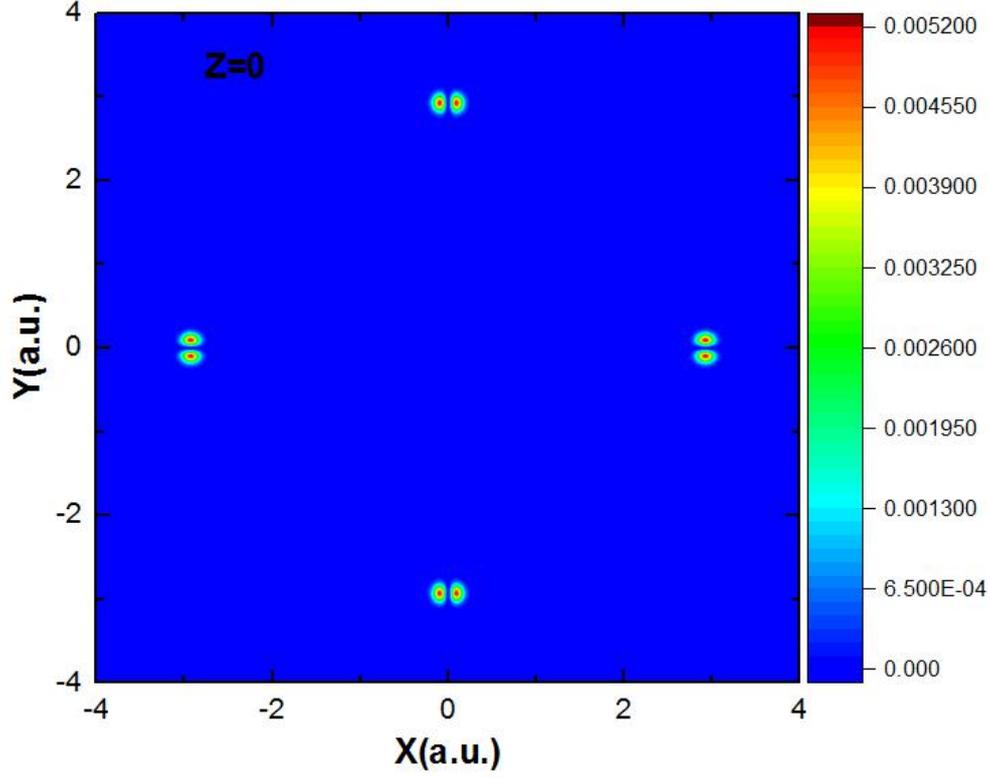

FIG. 4. The density distribution of the wave function by Eq. (14) in XY plane in coordinate space with the value of Z=0 a.u..

We now consider imaging the wave function of $\psi_2$ and $\psi_3$ by ATI spectra. Since the polarization direction of the two-color laser fields is along Z axis, the ATI spectra for both wave functions $\psi_2$ and $\psi_3$ are similar, hence we only present the results for wave function $\psi_2$.

The wave function of $\psi_2$ in momentum space is presented in Eq. (A5) of Appendix A, where the density distribution of this wave function is shown in Fig. 5.

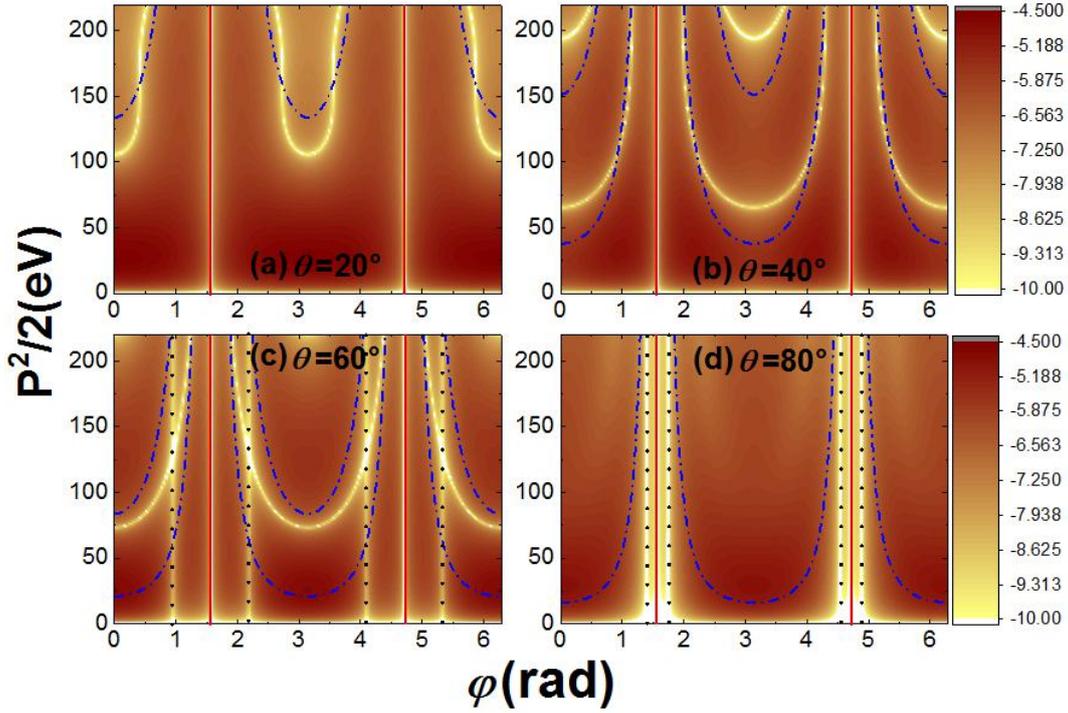

FIG. 5. The density distribution of $\psi_2$ wave function in momentum space for molecule SF$_6$.

Similarly for the case of $\psi_1$, there also exists three types of fringes: (1) when $\varphi=\pi/2+n\pi$ with n=1,2,…, the amplitude of the wave function is zero because each term in Eq. (A5) depends on the momentum p$_x$, which is denoted by the red solid line in Fig. 5; (2) when $\cos\varphi=\pm ctg\theta$, i.e., p$_x$=$\pm$ p$_z$, the interference between each two corresponding terms in Eq. (A5) provides the destructive fringes in the density distribution of the wave function, as shown by the black dotted lines in Figs. 5(c) and (d); (3) besides above fringes, we may also find some special fringes which carry the information of the molecular structure in the high energy region where these fringes can be predicted by the blue dash-dot lines in Fig. 5. These dash-dot lines can be obtained by simplifying the wave function of $\psi_2$ in Eq. (A5) with two terms for the momentum being large enough:

$$\psi_2 \approx -2.74966 \times 10^5 p_z \sin(R_{SF} p_x) e^{-\frac{p^2}{221.764}}$$
$$+ 2.74966 \times 10^5 p_x \sin(R_{SF} p_z) e^{-\frac{p^2}{221.764}} \quad (15)$$

Therefore, we may obtain an interference formula by the first term of above equation, which can be written as

$$E = 0.5(n\pi)^2 / (\sin^2 \theta_f \cos^2 \varphi_f R^2). \quad (16)$$

The blue dash-dot lines in Fig. 5 are drawn by Eq. (16). It can be found that these dash-dot lines agree with the fringes in high energy region, with $E$ larger than about 150eV, in the density distribution of the wave function.

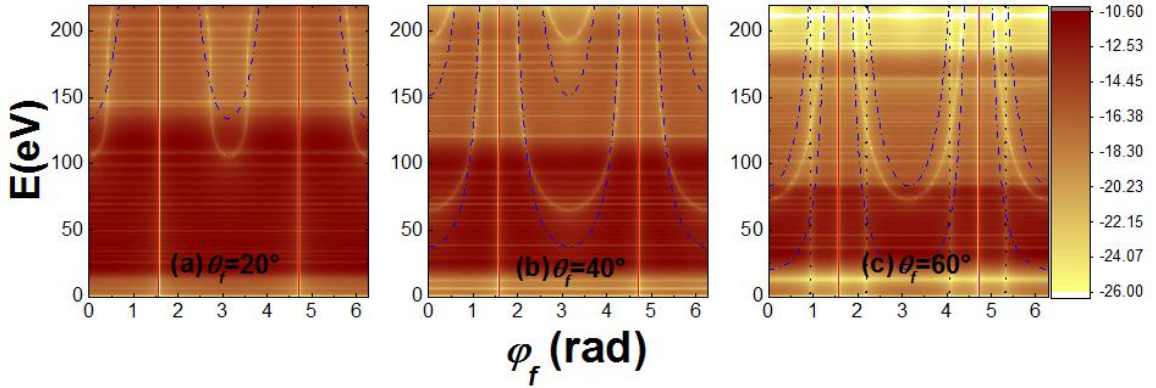

FIG. 6. The total angle-resolved ATI spectra of SF$_6$ for initial state $\psi_2$.

Figure 6 shows the angle-resolved ATI spectra of SF$_6$ for the initial state of $\psi_2$ with the angle $\theta_f = 20°$ (a), $40°$ (b) and $60°$ (c). The blue dash-dot lines in Fig. 6 are predicted by Eq. (16). By using these fringes, we can also derive the S-F bond length by the following formula, which is obtained from Eq. (16):

$$R = n\pi / (\sqrt{2E} |\sin \theta_f| |\cos \varphi_f|). \quad (17)$$

Table 3 presents the predictions of the S-F bond length by using the data of the numerical calculation shown in Fig. 6. We can find that the error shown in Table 3 is smaller than 10%.

Tab. 3. The calculation results of the S-F bond length of $SF_6$ when the symmetry is $\psi_2$ by Eq. (17).

| $\theta_f$ | $\varphi_f$ /rad | E /eV | R/a.u. | $R_{avg}$/a.u | Error$\Delta$ |
|---|---|---|---|---|---|
| **20°** | 0.42 | 190 | 2.69 | 2.69 | 8.16% |
|  | 2.72 | 190 | 2.69 |  |  |
|  | 3.56 | 190 | 2.69 |  |  |
|  | 5.87 | 190 | 2.68 |  |  |
| **40°** | 1.15 | 190 | 3.20 | 3.21 | 9.59% |
|  | 1.99 | 190 | 3.21 |  |  |
|  | 4.29 | 190 | 3.19 |  |  |
|  | 5.13 | 190 | 3.22 |  |  |
| **60°** | 1.19 | 190 | 2.70 | 2.68 | 8.50% |
|  | 1.95 | 190 | 2.71 |  |  |
|  | 4.32 | 190 | 2.63 |  |  |
|  | 5.10 | 190 | 2.66 |  |  |

Similarly, the wave function of Eq. (15) in momentum space corresponds to four effective terms of the wave function written as Eq. (18) in the coordinate space

$$\psi_2' = \frac{1}{\sqrt{2}} \{1.78005 z \exp\{-55.4441[(x-2.929)^2 + y^2 + z^2]\}$$
$$-1.78005 z \exp\{-55.4441[(x+2.929)^2 + y^2 + z^2]\}, \quad (18)$$
$$-1.78005 x \exp\{-55.4441[x^2 + y^2 + (z-2.929)^2]\}$$
$$+1.78005 x \exp\{-55.4441[x^2 + y^2 + (z+2.929)^2]\}$$

which is pertinent to four F atoms in XZ plane, as shows in Fig. 7.

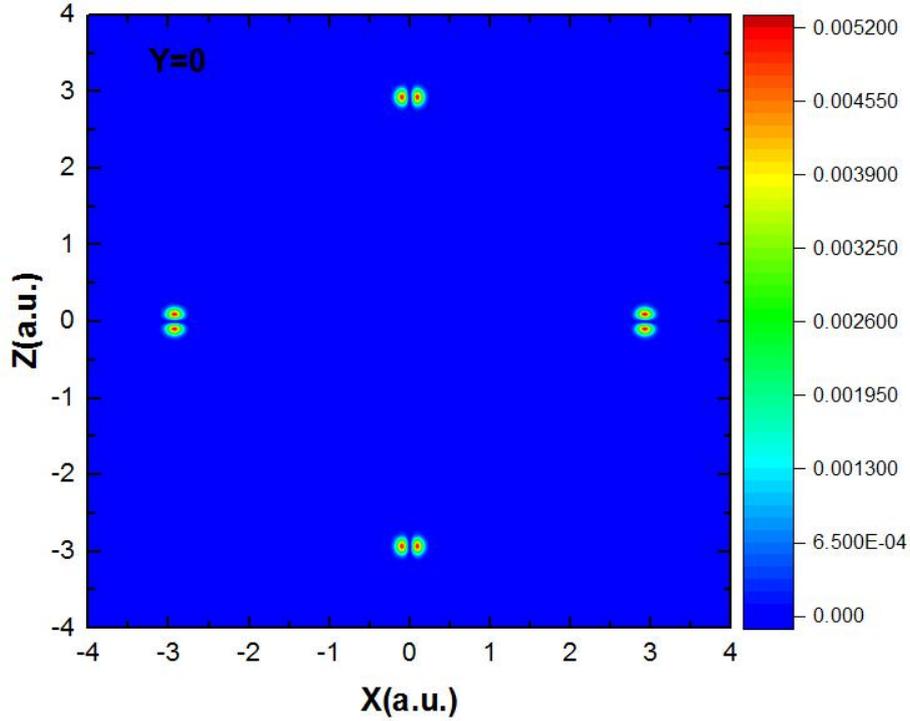

Fig. 7. The density distribution of the wave function Eq. (18) in XZ plane in coordinate space for Y=0 a.u..

Now we consider the ATI spectrum of the HOMO-1 orbits, where there exist three HOMO-1 degenerate orbitals denoted by $\psi_4$, $\psi_5$, and $\psi_6$. Fig. 8. shows the angle-resolved ATI spectra for the HOMO-1 wave function $\psi_4$, where we may find that the probability of this spectrum is much smaller than that of the corresponding ATI spectrum for the HOMO wave functions. The other two wave functions, i.e., $\psi_5$ and $\psi_6$, present similar results. One reason for these results is that the ionization threshold of HOMO-1 orbits, which is 14.6eV, is much smaller than XUV photon energy 77.5eV, thereby, although the detuning for HOMO-1 orbits is a little larger than that for HOMO orbits, this difference is so small that it can be ignored. The other crucially important reason is that the momentum space wave function of HOMO-1 orbits is more diffuse, which can be found in Eq. (B1) in the appendix B, thereby the ionization probability of the HOMO-1 orbits is much lower than that of the HOMO orbits for the photoelectron energy in the region of 0 to 220 eV. As a result, the contribution of the HOMO-1 wave functions to the ATI spectrum can be ignored under our present laser conditions.

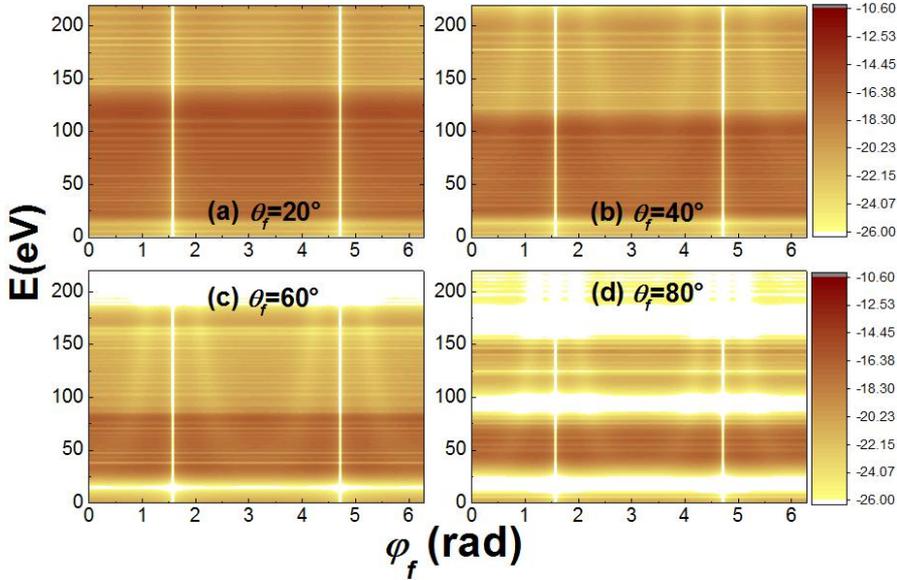

FIG. 8. The total angle-resolved ATI spectra of $SF_6$ for the HOMO-1 wave function $\psi_4$.

## IV. CONCLUSIONS

In summary, we have imaged the structures of complex symmetric polyatomic molecules $SF_6$ by using IR+XUV two-color laser fields. We find that the complex wave function of the molecule can be simplified by few terms in high energy region, hence some destructive interference formula carrying molecular information can be obtained by analyzing these terms. Furthermore, since the angle-resolved ATI spectrum is proportional to the molecular wave function, the destructive interference fringes in the ATI spectra in high energy region can be used to derive the molecular information according to the interference formulas which is from the simplified molecular wave function. For $SF_6$ molecule, because it has a high degree of symmetry, the destructive interference fringes in high energy part of ATI spectrum can provide effective information to predict the bond length of S-F for the three HOMO wave functions. In this work, we put forward a novel path of utilizing two color laser fields to extract the molecular structure information from frequency domain perspective, which is effective for imaging complex polyatomic molecules.

## ACKNOWLEDGEMENTS

This work was supported by the National Natural Science Foundation of China under Grant Nos. 11774411, 11474348, 11674198 and 11425414. And we thank all the members of strong-field atomic and molecular physics (SFAMP) club for their helpful discussions and suggestions.

# Appendix A

## $SF_6$

The molecular orbitals are from LCAO (linear combination of atomic orbitals), and atomic orbitals consist of Gaussian functions, and we can obtain three molecular orbitals in coordinate space in the following part. Firstly, the orbital of $\psi_1'$ can be written as,

$$\begin{aligned}
\psi_1' = \frac{1}{\sqrt{2}} \{ & 1.78005 y \exp\{-55.4441[(x-2.929)^2 + y^2 + z^2]\} + 1.87624 y \exp\{-12.6323[(x-2.929)^2 + y^2 + z^2]\} \\
& + 1.40521 y \exp\{-3.71756[(x-2.929)^2 + y^2 + z^2]\} - 1.78005 y \exp\{-55.4441[(x+2.929)^2 + y^2 + z^2]\} \\
& - 1.87624 y \exp\{-12.6323[(x+2.929)^2 + y^2 + z^2]\} - 1.40521 y \exp\{-3.71756[(x+2.929)^2 + y^2 + z^2]\} \\
& + 0.614912 y \exp\{-1.16545[(x-2.929)^2 + y^2 + z^2]\} - 0.614912 y \exp\{-1.16545[(x+2.929)^2 + y^2 + z^2]\} \\
& + 0.111126 y \exp\{-0.321892[(x-2.929)^2 + y^2 + z^2]\} - 0.111126 y \exp\{-0.321892[(x+2.929)^2 + y^2 + z^2]\} \\
& - 0.0863376(x-2.929) y \exp\{-1.75[(x-2.929)^2 + y^2 + z^2]\} - 0.0863376(x+2.929) y \exp\{-1.75[(x+2.929)^2 + y^2 + z^2]\} \\
& - 1.78005 x \exp\{-55.4441[x^2 + (y-2.929)^2 + z^2]\} - 1.87624 x \exp\{-12.6323[x^2 + (y-2.929)^2 + z^2]\} \\
& - 1.40521 x \exp\{-3.71756[x^2 + (y-2.929)^2 + z^2]\} + 1.78005 x \exp\{-55.4441[x^2 + (y+2.929)^2 + z^2]\} \\
& + 1.87624 x \exp\{-12.6323[x^2 + (y+2.929)^2 + z^2]\} + 1.40521 x \exp\{-3.71756[x^2 + (y+2.929)^2 + z^2]\} \\
& - 0.614912 x \exp\{-1.16545[x^2 + (y-2.929)^2 + z^2]\} + 0.614912 x \exp\{-1.16545[x^2 + (y-2.929)^2 + z^2]\} \\
& - 0.111126 x \exp\{-0.321892[x^2 + (y-2.929)^2 + z^2]\} + 0.111126 x \exp\{-0.321892[x^2 + (y+2.929)^2 + z^2]\} \\
& + 0.0863376 x (y-2.929) \exp\{-1.75[x^2 + (y-2.929)^2 + z^2]\} + 0.0863376 x (y+2.929) \exp\{-1.75[x^2 + (y+2.929)^2 + z^2]\} \}
\end{aligned}$$
,(A1)

which can be depicted in Fig. A1.

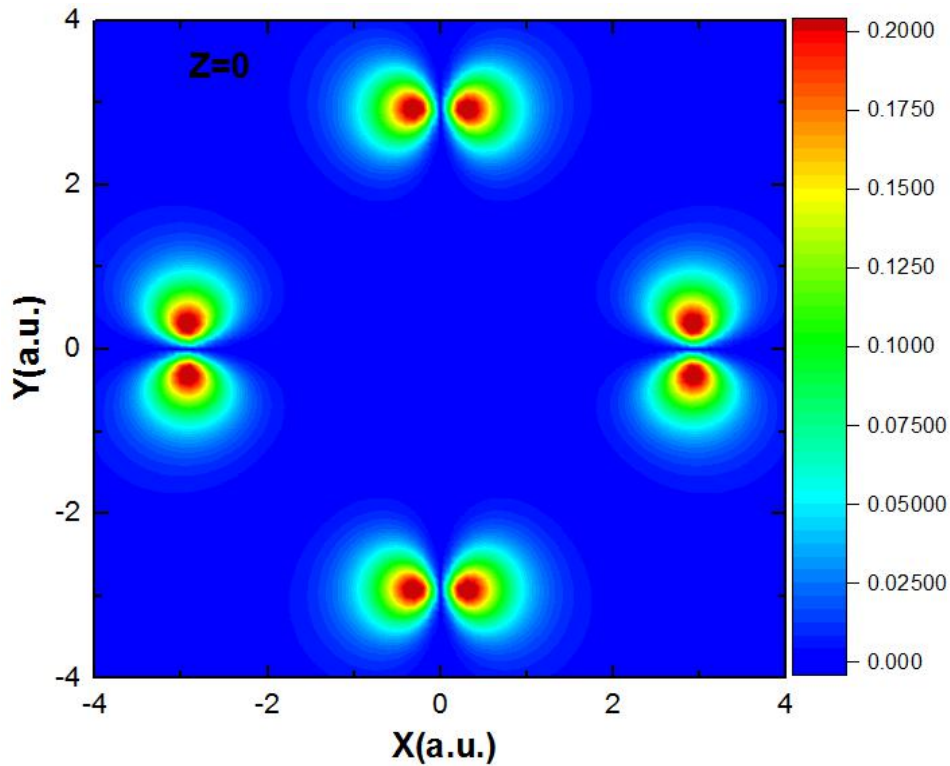

FIG. A1. The density distribution of wave function $\psi_1'$ in coordinate space with the value of Z=0 a.u..

The wave function for the symmetry of $\psi_2'$ is

$$\psi'_2 = \frac{1}{\sqrt{2}}\{1.78005z\exp\{-55.4441[(x-2.929)^2+y^2+z^2]\}+1.87624z\exp\{-12.6323[(x-2.929)^2+y^2+z^2]\}$$
$$+1.40521z\exp\{-3.71756[(x-2.929)^2+y^2+z^2]\}-1.78005z\exp\{-55.4441[(x+2.929)^2+y^2+z^2]\}$$
$$-1.87624z\exp\{-12.6323[(x+2.929)^2+y^2+z^2]\}-1.40521z\exp\{-3.71756[(x+2.929)^2+y^2+z^2]\}$$
$$+0.614912z\exp\{-1.16545[(x-2.929)^2+y^2+z^2]\}-0.614912z\exp\{-1.16545[(x+2.929)^2+y^2+z^2]\}$$
$$+0.111126z\exp\{-0.321892[(x-2.929)^2+y^2+z^2]\}-0.111126z\exp\{-0.321892[(x+2.929)^2+y^2+z^2]\}$$
$$-0.0863376(x-2.929)z\exp\{-1.75[(x-2.929)^2+y^2+z^2]\}-0.0863376(x+2.929)z\exp\{-1.75[(x+2.929)^2+y^2+z^2]\}$$
$$-1.78005x\exp\{-55.4441[x^2+y^2+(z-2.929)^2]\}-1.87624x\exp\{-12.6323[x^2+y^2+(z-2.929)^2]\}$$
$$-1.40521x\exp\{-3.71756[x^2+y^2+(z-2.929)^2]\}+1.78005x\exp\{-55.4441[x^2+y^2+(z+2.929)^2]\}$$
$$+1.87624x\exp\{-12.6323[x^2+y^2+(z+2.929)^2]\}+1.40521z\exp\{-3.71756[x^2+y^2+(z+2.929)^2]\}$$
$$-0.614912x\exp\{-1.16545[x^2+y^2+(z-2.929)^2]\}+0.614912x\exp\{-1.16545[x^2+y^2+(z+2.929)^2]\}$$
$$-0.111126x\exp\{-0.321892[x^2+y^2+(z-2.929)^2]\}+0.111126x\exp\{-0.321892[x^2+y^2+(z+2.929)^2]\}$$
$$+0.0863376x(z-2.929)\exp\{-1.75[x^2+y^2+(z-2.929)^2]\}+0.0863376x(z+2.929)\exp\{-1.75[x^2+y^2+(z+2.929)^2]\}\}$$

.(A2)

which is shown by Fig. A2.

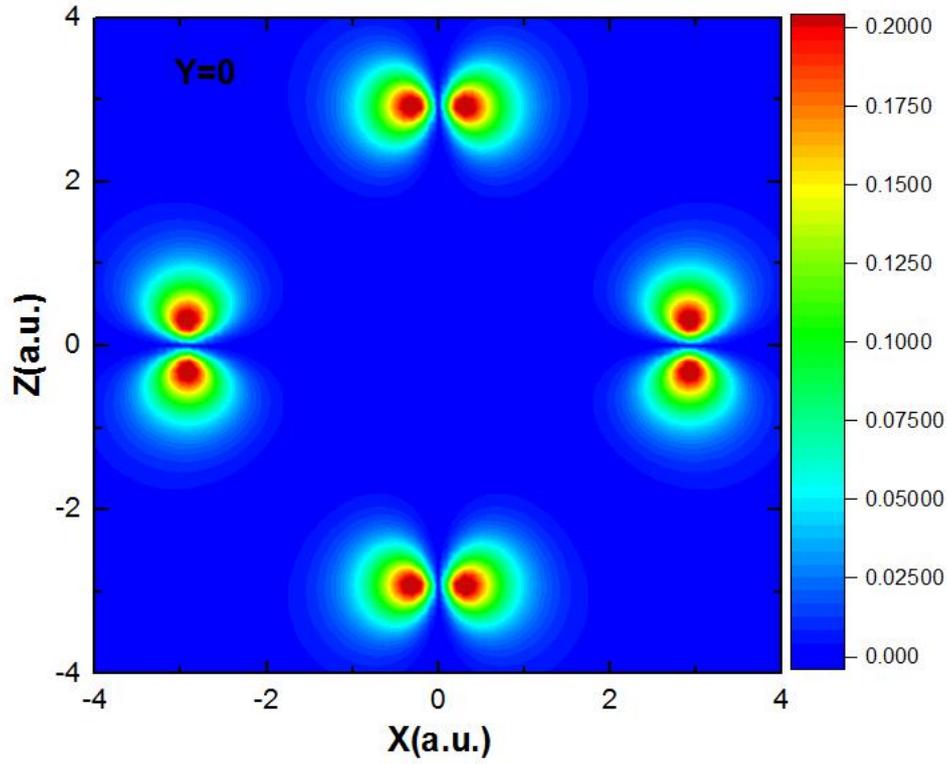

FIG. A2. The density distribution of wave function $\psi'_2$ in coordinate space with the value of Y=0 a.u..

The coordinate space wave function of $\psi'_3$ is

$$\psi'_3 = \frac{1}{\sqrt{2}}\{1.78005z\exp\{-55.4441[x^2+(y-2.929)^2+z^2]\}+1.87624z\exp\{-12.6323[x^2+(y-2.929)^2+z^2]\}$$
$$+1.40521z\exp\{-3.71756[x^2+(y-2.929)^2+z^2]\}-1.78005z\exp\{-55.4441[x^2+(y+2.929)^2+z^2]\}$$
$$-1.87624z\exp\{-12.6323[x^2+(y+2.929)^2+z^2]\}-1.40521z\exp\{-3.71756[x^2+(y+2.929)^2+z^2]\}$$
$$+0.614912z\exp\{-1.16545[x^2+(y-2.929)^2+z^2]\}-0.614912z\exp\{-1.16545[x^2+(y+2.929)^2+z^2]\}$$
$$+0.111126z\exp\{-0.321892[x^2+(y-2.929)^2+z^2]\}-0.111126z\exp\{-0.321892[x^2+(y+2.929)^2+z^2]\}$$
$$-0.0863376(y-2.929)z\exp\{-1.75[x^2+(y-2.929)^2+z^2]\}-0.0863376(y+2.929)z\exp\{-1.75[x^2+(y+2.929)^2+z^2]\}$$
$$-1.78005y\exp\{-55.4441[x^2+y^2+(z-2.929)^2]\}-1.87624y\exp\{-12.6323[x^2+y^2+(z-2.929)^2]\}$$
$$-1.40521y\exp\{-3.71756[x^2+y^2+(z-2.929)^2]\}+1.78005y\exp\{-55.4441[x^2+y^2+(z+2.929)^2]\}$$
$$+1.87624y\exp\{-12.6323[x^2+y^2+(z+2.929)^2]\}+1.40521y\exp\{-3.71756[x^2+y^2+(z+2.929)^2]\}$$
$$-0.614912y\exp\{-1.16545[x^2+y^2+(z-2.929)^2]\}+0.614912y\exp\{-1.16545[x^2+y^2+(z+2.929)^2]\}$$
$$-0.111126y\exp\{-0.321892[x^2+y^2+(z-2.929)^2]\}+0.111126y\exp\{-0.321892[x^2+y^2+(z+2.929)^2]\}\}$$
$$+0.0863376y(z-2.929)\exp\{-1.75[x^2+y^2+(z-2.929)^2]\}+0.0863376y(z+2.929)\exp\{-1.75[x^2+y^2+(z+2.929)^2]\}$$

.(A3)

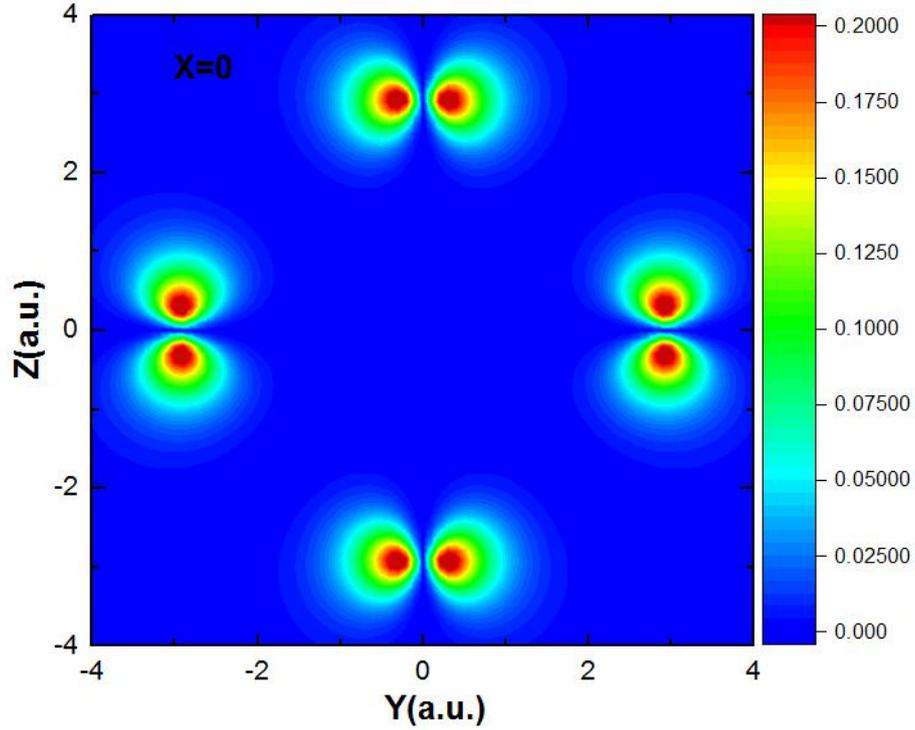

FIG. A3. The density distribution of wave function $\psi'_3$ with the value of X=0 a.u..

By the Fourier transform, the wave functions of $\psi'_1$, $\psi'_2$ and $\psi'_3$ can be expressed in momentum space by $\psi_1$, $\psi_2$ and $\psi_3$, separately, which is

$$\psi_1 = \frac{1}{\sqrt{6482.2115}}[-0.0000274966p_y \sin(R_{SF}p_x)e^{-\frac{p^2}{221.764}} - 0.0011696p_y \sin(R_{SF}p_x)e^{-\frac{p^2}{50.5292}}$$

$$-0.0186445p_y \sin(R_{SF}p_x)e^{-\frac{p^2}{14.8702}} - 0.148263p_y \sin(R_{SF}p_x)e^{-\frac{p^2}{4.6618}}$$

$$-0.668336p_y \sin(R_{SF}p_x)e^{-\frac{p^2}{1.28757}} + 0.00215274p_xp_y \cos(R_{SF}p_x)e^{-\frac{p^2}{7}}$$

$$+0.0000274966p_x \sin(R_{SF}p_y)e^{-\frac{p^2}{221.764}} + 0.0011696p_x \sin(R_{SF}p_y)e^{-\frac{p^2}{50.5292}}$$

$$+0.0186445p_x \sin(R_{SF}p_y)e^{-\frac{p^2}{14.8702}} + 0.148263p_x \sin(R_{SF}p_y)e^{-\frac{p^2}{4.6618}}$$

$$+0.668336p_x \sin(R_{SF}p_y)e^{-\frac{p^2}{1.28757}} - 0.00215274p_xp_y \cos(R_{SF}p_y)e^{-\frac{p^2}{7}}], \quad (A4)$$

$$\psi_2 = \frac{1}{\sqrt{6565.512}}[-0.0000274966p_z \sin(R_{SF}p_x)e^{-\frac{p^2}{221.764}} - 0.0011696p_z \sin(R_{SF}p_x)e^{-\frac{p^2}{50.5292}}$$

$$-0.0186445p_z \sin(R_{SF}p_x)e^{-\frac{p^2}{14.8702}} - 0.148263p_z \sin(R_{SF}p_x)e^{-\frac{p^2}{4.6618}}$$

$$-0.668336p_z \sin(R_{SF}p_x)e^{-\frac{p^2}{1.28757}} + 0.00215274p_xp_z \cos(R_{SF}p_x)e^{-\frac{p^2}{7}}$$

$$+0.0000274966p_x \sin(R_{SF}p_z)e^{-\frac{p^2}{221.764}} + 0.0011696p_x \sin(R_{SF}p_z)e^{-\frac{p^2}{50.5292}}$$

$$+0.0186445p_x \sin(R_{SF}p_z)e^{-\frac{p^2}{14.8702}} + 0.148263p_x \sin(R_{SF}p_z)e^{-\frac{p^2}{4.6618}}$$

$$+0.668336p_x \sin(R_{SF}p_z)e^{-\frac{p^2}{1.28757}} - 0.00215274p_xp_z \cos(R_{SF}p_z)e^{-\frac{p^2}{7}}], \quad (A5)$$

and

$$\psi_3 = \frac{1}{\sqrt{6565.512}}[-0.0000274966p_z \sin(R_{SF}p_y)e^{-\frac{p^2}{221.764}} - 0.0011696p_z \sin(R_{SF}p_y)e^{-\frac{p^2}{50.5292}}$$

$$-0.0186445p_z \sin(R_{SF}p_y)e^{-\frac{p^2}{14.8702}} - 0.148263p_z \sin(R_{SF}p_y)e^{-\frac{p^2}{4.6618}}$$

$$-0.668336p_z \sin(R_{SF}p_y)e^{-\frac{p^2}{1.28757}} + 0.00215274p_yp_z \cos(R_{SF}p_y)e^{-\frac{p^2}{7}}$$

$$+0.0000274966p_y \sin(R_{SF}p_z)e^{-\frac{p^2}{221.764}} + 0.0011696*p_y \sin(R_{SF}p_z)e^{-\frac{p^2}{50.5292}}$$

$$+0.0186445p_y \sin(R_{SF}p_z)e^{-\frac{p^2}{14.8702}} + 0.148263p_y \sin(R_{SF}p_z)e^{-\frac{p^2}{4.6618}}$$

$$+0.668336p_y \sin(R_{SF}p_z)e^{-\frac{p^2}{1.28757}} - 0.00215274p_yp_z \cos(R_{SF}p_z)e^{-\frac{p^2}{7}}]. \quad (A6)$$

## Appendix B

We can obtain three wave functions for the HOMO-1 orbitals of $SF_6$ in momentum space in the following part, which is expressed as

$$\psi_4 = \frac{1}{\sqrt{6.0649141 \times 10^{12}}} \{-ip_x(2.03526 \times 10^{-8} e^{-\frac{p^2}{1980.16}} + 9.4942 \times 10^{-7} e^{-\frac{p^2}{468.884}} + 0.0000169554 e^{-\frac{p^2}{151.1}}$$

$$+0.00015655 e^{-\frac{p^2}{56.2336}} + 0.00093331 e^{-\frac{p^2}{22.263}} + 0.0020788 e^{-\frac{p^2}{9.05188}}$$

$$-0.0354173 e^{-\frac{p^2}{3.23198}} - 0.0175772 e^{-\frac{p^2}{1.10984}} + 0.467701 e^{-\frac{p^2}{0.308564}})$$

$$+i\sin(p_x R_{SF})(1.24989 \times 10^{-9} e^{-\frac{p^2}{45708.4}} + 3.94247 \times 10^{-8} e^{-\frac{p^2}{6889.4}}$$

$$+5.88987 \times 10^{-7} e^{-\frac{p^2}{1582.98}} + 5.09161 \times 10^{-6} e^{-\frac{p^2}{460.556}} + 0.0000323751 e^{-\frac{p^2}{134.41}}$$

$$+0.0000695514 e^{-\frac{p^2}{19.676}} + 8.20719 \times 10^{-6} e^{-\frac{p^2}{221.776}} + 0.000200014 e^{-\frac{p^2}{50.5292}}$$

$$-1.837 \times 10^{-6} e^{-\frac{p^2}{14.8702}} - 0.0110525 e^{-\frac{p^2}{4.6618}} + 0.0826785 e^{-\frac{p^2}{1.28757}})$$

$$-ip_x \cos(p_x R_{SF})(0.0000178434 e^{-\frac{p^2}{221.776}} + 0.00075904 e^{-\frac{p^2}{50.5292}}$$

$$+0.0120998 e^{-\frac{p^2}{14.8702}} + 0.104527 e^{-\frac{p^2}{4.6618}} + 0.487334 e^{-\frac{p^2}{1.28757}})$$

$$+i\sin(p_x R_{SF}) e^{-\frac{p^2}{7}} [-0.000569921(p_x^2 + p_y^2 - 2p_z^2) + 0.00170975(p_y^2 - p_x^2)]$$

$$-ip_x \cos(p_y R_{SF})(0.0000226208 e^{-\frac{p^2}{221.776}} + 0.000962267 e^{-\frac{p^2}{50.5292}} + 0.0153394 e^{-\frac{p^2}{14.8702}}$$

$$+0.121283 e^{-\frac{p^2}{4.6618}} + 0.543345 e^{-\frac{p^2}{1.28757}}) - 0.00167446 ip_x p_y \sin(p_y R_{SF}) e^{-\frac{p^2}{7}}$$

$$-ip_x \cos(p_z R_{SF})(0.0000226208 e^{-\frac{p^2}{221.776}} + 0.000962267 e^{-\frac{p^2}{50.5292}} + 0.0153394 e^{-\frac{p^2}{14.8702}}$$

$$+0.121283 e^{-\frac{p^2}{4.6618}} + 0.543345 e^{-\frac{p^2}{1.28757}} - 0.00167446 ip_x p_z \sin(p_z R_{SF}) e^{-\frac{p^2}{7}} \}$$

, (B1)

$$\psi_5 = \frac{1}{\sqrt{6.0646892 \times 10^{12}}} \{-ip_y(2.03526 \times 10^{-8} e^{-\frac{p^2}{1980.16}} + 9.4942 \times 10^{-7} e^{-\frac{p^2}{468.884}} + 0.0000169554 e^{-\frac{p^2}{151.1}}$$

$$+0.00015655 e^{-\frac{p^2}{56.2336}} + 0.00093331 e^{-\frac{p^2}{22.263}} + 0.00207884 e^{-\frac{p^2}{9.05188}}$$

$$-0.0354173 e^{-\frac{p^2}{3.23198}} - 0.0175772 e^{-\frac{p^2}{1.10984}} + 0.467702 e^{-\frac{p^2}{0.308564}})$$

$$+ip_y \cos(p_x R_{SF})(-0.0000226208 e^{-\frac{p^2}{221.776}} + 0.000962267 e^{-\frac{p^2}{50.5292}}$$

$$+0.0153394 e^{-\frac{p^2}{14.8702}} - 0.121283 e^{-\frac{p^2}{4.6618}} - 0.543345 e^{-\frac{p^2}{1.28757}})$$

$$-0.00221509 i \sin(p_x R_{SF}) p_x p_y e^{-\frac{p^2}{7}}$$

$$+i \sin(p_y R_{SF})(1.24989 \times 10^{-9} e^{-\frac{p^2}{45708.4}} + 3.94247 \times 10^{-8} e^{-\frac{p^2}{6889.4}}$$

$$+5.88987 \times 10^{-7} e^{-\frac{p^2}{1582.98}} + 5.09161 \times 10^{-6} e^{-\frac{p^2}{460.556}} + 0.0000323751 e^{-\frac{p^2}{134.41}}$$

$$+0.0000695514 e^{-\frac{p^2}{19.676}} + 8.20719 \times 10^{-6} e^{-\frac{p^2}{221.776}} + 0.000200014^{-\frac{p^2}{50.5292}}$$

$$-1.837 \times 10^{-6} e^{-\frac{p^2}{14.8702}} - 0.0110525 e^{-\frac{p^2}{4.6618}} + 0.0826785 e^{-\frac{p^2}{1.28757}})$$

$$-ip_y \cos(p_y R_{SF})(0.0000178434 e^{-\frac{p^2}{221.776}} + 0.000759064 e^{-\frac{p^2}{50.528}}$$

$$+0.0120998 e^{-\frac{p^2}{14.8702}} + 0.104527 e^{-\frac{p^2}{4.6618}} + 0.487334 e^{-\frac{p^2}{1.28757}})$$

$$-i \sin(p_y R_{SF})(p_x^2 + p_y^2 - 2p_z^2) 0.000569921 e^{-\frac{p^2}{7}}$$

$$-i \sin(p_y R_{SF})(p_y^2 - p_x^2) 0.00170976 e^{-\frac{p^2}{7}}$$

$$-ip_y \cos(p_z R_{SF})(0.0000226208 e^{-\frac{p^2}{221.776}} + 0.000962267 e^{-\frac{p^2}{50.5292}}$$

$$+0.0153394 e^{-\frac{p^2}{14.8702}} + 0.121283 e^{-\frac{p^2}{4.6618}} + 0.543345 e^{-\frac{p^2}{1.28757}})$$

$$-ip_y p_z \sin(p_z R_{SF}) 0.00167446 e^{-\frac{p^2}{7}} \}$$

, (B2)

and

$$\psi_6 = \frac{1}{\sqrt{6.0646707 \times 10^{12}}} \{-ip_z (2.03526 \times 10^{-8} e^{-\frac{p^2}{1980.16}} + 9.4942 \times 10^{-7} e^{-\frac{p^2}{468.884}} + 0.0000169554 e^{-\frac{p^2}{151.1}}$$

$$+ 0.00015655 e^{-\frac{p^2}{56.2336}} + 0.00093331 e^{-\frac{p^2}{22.263}} + 0.00207884 e^{-\frac{p^2}{9.05188}}$$

$$- 0.0354173 e^{-\frac{p^2}{3.23198}} - 0.0175772 e^{-\frac{p^2}{1.10984}} + 0.467702 e^{-\frac{p^2}{0.308564}})$$

$$- ip_z \cos(p_x R_{SF})(0.0000226208 e^{-\frac{p^2}{221.776}} + 0.000962267 e^{-\frac{p^2}{50.5292}}$$

$$+ 0.0153394 e^{-\frac{p^2}{14.8702}} + 0.12183 e^{-\frac{p^2}{4.6618}} + 0.543345 e^{-\frac{p^2}{1.28757}})$$

$$- i \sin(p_x R_{SF}) p_x p_z 0.00167446 e^{-\frac{p^2}{7}}$$

$$- ip_z \cos(p_y R_{SF})(0.0000226208 e^{-\frac{p^2}{221.776}} + 0.000962267 e^{-\frac{p^2}{50.5292}}$$

$$+ 0.0153394 e^{-\frac{p^2}{14.8702}} + 0.121283 e^{-\frac{p^2}{4.6618}} + 0.543345 e^{-\frac{p^2}{1.28757}})$$

$$- ip_y p_z \sin(p_y R_{SF}) 0.00167446 e^{-\frac{p^2}{7}}$$

$$+ i \sin(p_z R_{SF})(1.24989 \times 10^{-9} e^{-\frac{p^2}{45708.4}} + 3.94247 \times 10^{-8} e^{-\frac{p^2}{6889.4}}$$

$$+ 5.88987 \times 10^{-7} e^{-\frac{p^2}{1582.98}} + 5.09161 \times 10^{-6} e^{-\frac{p^2}{460.556}} + 0.0000323751 e^{-\frac{p^2}{134.41}}$$

$$+ 0.0000695514 e^{-\frac{p^2}{19.676}} + 8.20719 \times 10^{-6} e^{-\frac{p^2}{221.776}} + 0.000200014 e^{-\frac{p^2}{50.5292}}$$

$$- 1.837 \times 10^{-6} e^{-\frac{p^2}{14.8702}} - 0.0110525 e^{-\frac{p^2}{4.6618}} + 0.0826785 e^{-\frac{p^2}{1.28757}})$$

$$- ip_z \cos(p_z R_{SF})(0.0000178434 e^{-\frac{p^2}{221.776}} + 0.00075904 e^{-\frac{p^2}{50.5292}}$$

$$+ 0.0120998 e^{-\frac{p^2}{14.8702}} + 0.104527 e^{-\frac{p^2}{4.6618}} + 0.487334 e^{-\frac{p^2}{1.28757}})$$

$$+ i \sin(p_z R_{SF})(p_x^2 + p_y^2 - 2p_z^2) 0.00113979 e^{-\frac{p^2}{7}}\}$$

. (B3)